\documentclass[11pt,twoside]{article}

\usepackage{prcsa2008}
\usepackage[dvips]{graphicx}

\usepackage{natbib}

\begin{document}
\title{The ``Living with a Red Dwarf'' Program}   
\author{E.F.~Guinan and S.G.~Engle}   
\affil{Villanova University, 800 E Lancaster Ave, Villanova, PA 19085, USA}    

\begin{abstract}
Red Dwarfs (main-sequence / dwarf M or dM) stars are the most common stars
in the Galaxy. These cool, faint, low mass stars comprise over 75\% of all
stars.  Because of their low luminosities ($\sim$0.0008--0.06 of the Sun's
luminosity), the circumstellar habitable zones (HZs) of dM stars are located
within $\sim$0.05--0.4 AU of the host star. Nevertheless, the prospect of
life on a planet located within the HZ of a red dwarf is moderately high,
based on the longevity of these stars ($>$50 Gyr), their constant
luminosities and high space densities.  Here we describe the aims and
early results of the ``Living with a Red Dwarf'' Program -- a study of dM
stars that we have been carrying out over the last few years. The primary
focus of our research on dM stars is the study of their magnetic dynamos
and resulting star spots \& coronal X-ray and chromospheric UV emissions
as a function of age, rotation and spectral type. This program will
provide datasets that can  be used as inputs for the study of all aspects
of dM stars, along with the planets already discovered hosted by them and
the probable hundreds (thousands?) of planets expected to be uncovered in
the near future by missions such as Kepler \& Darwin/TPF. These datasets
will be invaluable to those who model exo-planetary atmospheres, as well
as exobiologists \& astrobiologists who are studying the possibilities of
life elsewhere in the universe.  Our expected results will provide
fundamental information on the most numerous stars in the Galaxy --
including their ages, as well as radiative (irradiances) and magnetic
dynamo properties.  A significant element of our program is the determination
of accurate stellar magnetic-driven X-ray--UV (X-UV) irradiances that
are generated by the dM stars' vigorous magnetic dynamos. These X-UV
irradiances (and flare frequencies) are strongly dependent on rotation,
and thus age, and diminish as the stars lose angular momentum and
spin-down over time via magnetic braking.
\end{abstract}

\section{Background \& Introduction}

Red dwarf stars (also known as main-sequence M or dwarf M -- dM stars) are
by far the most numerous stars in our Galaxy, comprising more than 75\% of
all stars (see Fig. 1a). dM stars are cool, low luminosity stars with
deep convective zones and luminosities that range from
L $\approx$ 0.0008--0.06L$_{\odot}$ (for dM8--dM0 stars, respectively).
These diminutive low mass stars ($\sim$0.1--0.6 R$_{\odot}$;
$\sim$0.1--0.6 M$_{\odot}$) have very slow nuclear fusion rates and thus
very long lifetimes that range from $\sim$40 Gyr for dM0 stars to
$>$200 Gyr for the lower mass, very low luminosity dM5--8 stars.
Because of their long stable lifetimes, it is possible that planets hosted
by older dM stars could harbor life -- possibly even advanced intelligent
life. For example, the Sloan Digital Sky Survey has found millions of dM
stars as old as, and older than, the Sun \citep{bochan2007}. Given their
large numbers and long lifetimes, determining the number of dM stars with
planets and assessing planetary habitability is critically important
because such studies would indicate how common life is in the universe.

\begin{figure}
\begin{center}
\includegraphics[width=0.45\textwidth]{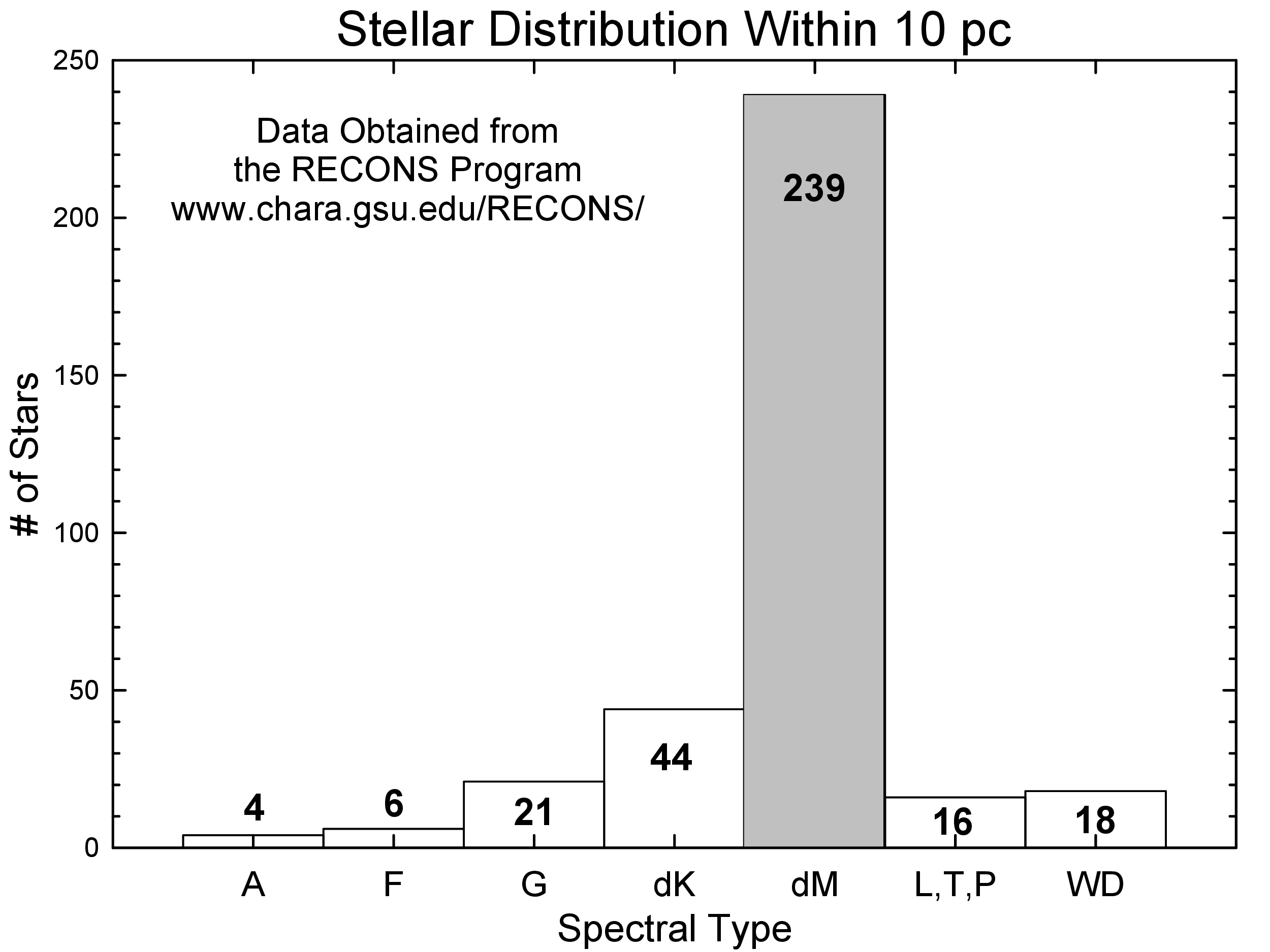}
\includegraphics[width=0.45\textwidth]{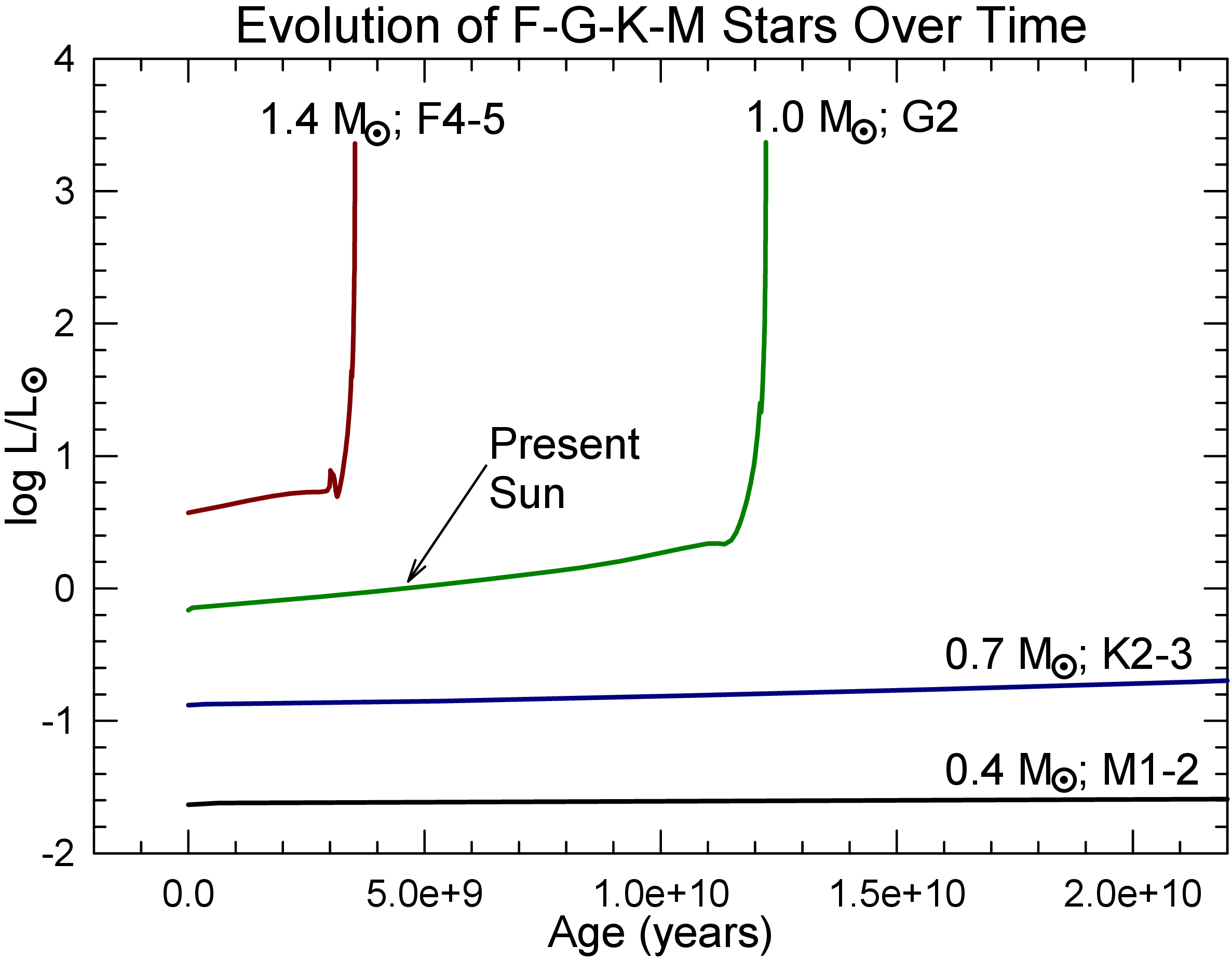}
\end{center}
\vspace{-5mm}
\caption{{\bf Left --} Histogram of the nearby stars (within 10 pc)
distributed according to spectral type bins. The data is from the RECONS
Program. Note dM stars represent over 75\% of the main-sequence stars.
{\bf Right --} The changes in the luminosity over time for representative
F5V, G2V, K2V, and M1--2V stars. The evolution calculations are from BaSTI
(http://albione.oa-teramo.inaf.it/).  Note that the luminosities of the
lower mass dK and dM stars change very slowly with time.}
\vspace{-5mm}
\end{figure}

Here we present the initial results of our exploratory pilot program
investigating photometry, spectroscopy, and space X-UV observations
for a representative sample of $\sim$40 dM0--6 stars. From this study we
have found strong evidence for relations among age, rotation and magnetic
activity tracers such as (coronal) X-ray, (transition region /
chromospheric) FUV--UV and (chromospheric) H$\alpha$ emission
\citep{guinanengle2007,guinan2007}.  Utilizing available survey photometry,
along with our own, we have identified rotation periods for over 50 dM
stars from star spot induced periodic, low amplitude light variations. We
have used the ROSAT X-ray and IUE UV archival data and published magnetic
field fluxes ($Bf$) to investigate the X-UV irradiances and their
correlations to magnetic field strengths (from \citet{reinersbasri2007} and
others). This preliminary study has resulted in promising Age-Rotation-Activity
relationships for dM0--6 stars. For excellent reviews of the current state
of research on -- dM star activity and evolution; their suitability as
planetary hosts; their impacts on possibly habitable planets and the current
planetary search programs including dM star targets -- see \citet{scalo2007,tarter2007} and references therein. Information
about the ``Living with a Red Dwarf'' Program can be found at
http://www.astronomy.villanova.edu/lward/. 

\section{Motivations for the Program}

\subsection{Stellar Habitable Zones (HZs)}

The temperature of a planet and the circumstellar liquid water HZs for
Earth-like planets depend strongly on the luminosity of the host star and
the planet's distance (1/$d^2$) from the star (e.g.,
\citet{kastingcatling2003}). Also important are the planet's albedo($A$), cloudiness, greenhouse
gas heat trapping contributions and the spectral energy distribution of
the host star \citep[see][]{selsis2007}. A relation for estimating the
average equilibrium temperature $T_{\rm P}$ of a rapidly rotating planet
(P$_{\rm rot} \leq$ 3--5 days) 
located at a distance ($d$) from the host star with a luminosity of $L$
(in solar units) is given by equation (1). 

\begin{equation}
{T_{\rm P}(K)} = \frac{279 [(1-A) L_*/L_\odot]^{1/4}}{d^{1/2}} + \Delta T_{\rm GH}
\end{equation}

Where  $A$ $=$ albedo, $L_*/L_\odot$ $=$ bolometric luminosity of the
star relative to the Sun and $d$ $=$ average distance of the hosted
planet from the star in Astronomical Units (1AU = 1.496 x 10$^8$
km) and $\Delta T_{\rm GH}$ $=$ Greenhouse Effect temperature enhancement
(in units of degrees K).  For the Earth and Sun: $d$ = 1.0 AU, $L_*/L_\odot$
= 1.0, $A$ $\approx$ 0.3, and $\Delta T_{\rm GH}$ $\sim$ 31 K, resulting in the
observed average equilibrium temperature of $T_\oplus$ $\approx$ 286K
( = +13$\deg$C = +55$\deg$F). Because of the
low luminosities of dM stars, their HZs are located very close to the
central star ($<$0.4 AU). Accordingly, a dM/HZ planet will be more strongly
influenced
by stellar flares, winds and coronal plasma ejection events that are
frequent in young dM stars (e.g. \citet{kasting1993,lammer2007}).
As comparison, for an Earth-like planet with an albedo and greenhouse
effect similar to the Earth, and hosted by a $\sim$dM2--3 star (with
L$_*$/L$_\odot$
$\sim$ 0.02) to have the same average temperature of our Earth, it would have
be located $d$ $\approx$ 0.14 AU from its host star. This is referred to
as the Earth-equivalent distance.

\begin{figure}[!t]
\begin{center}
\includegraphics[width=0.48\textwidth]{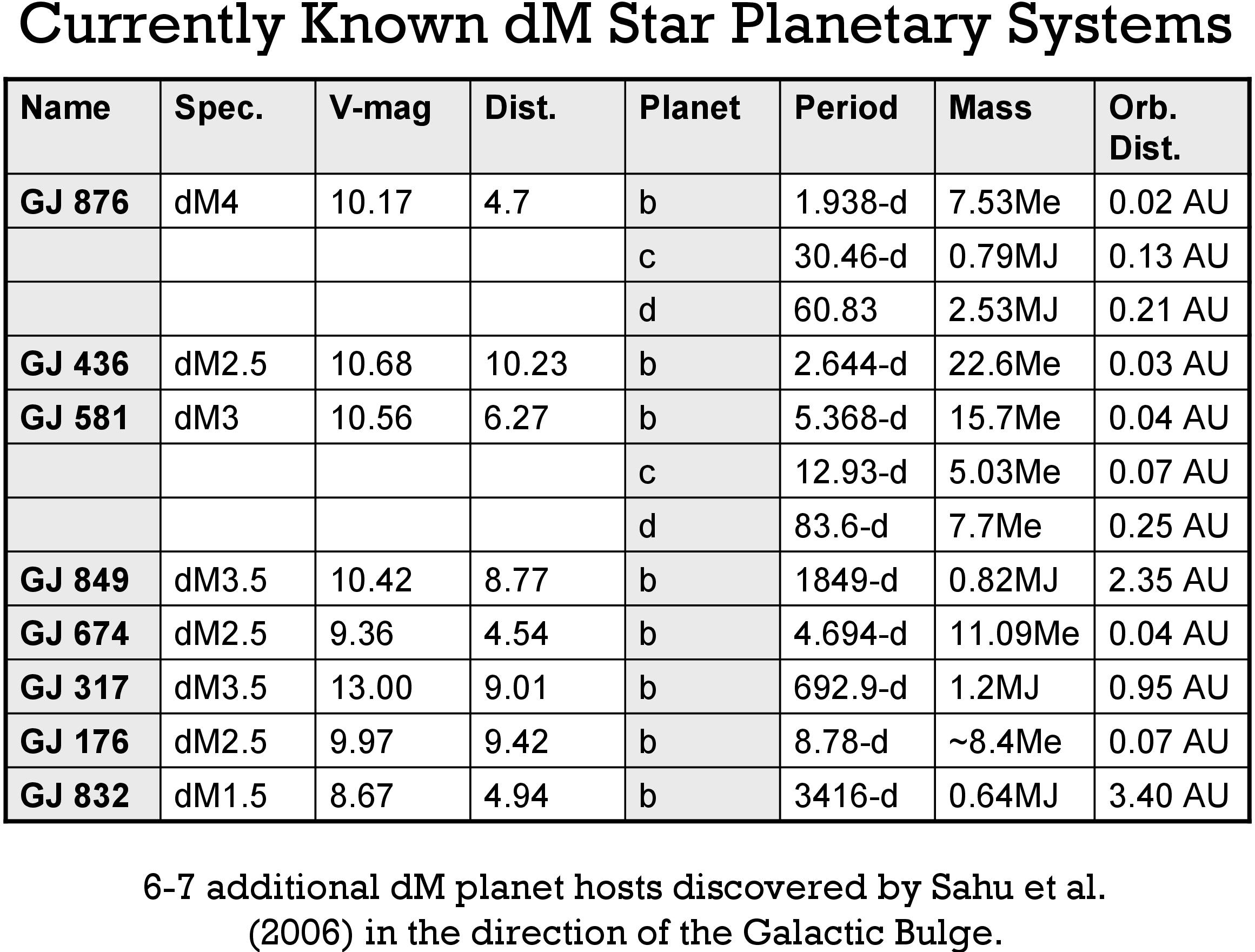}
\includegraphics[width=0.48\textwidth]{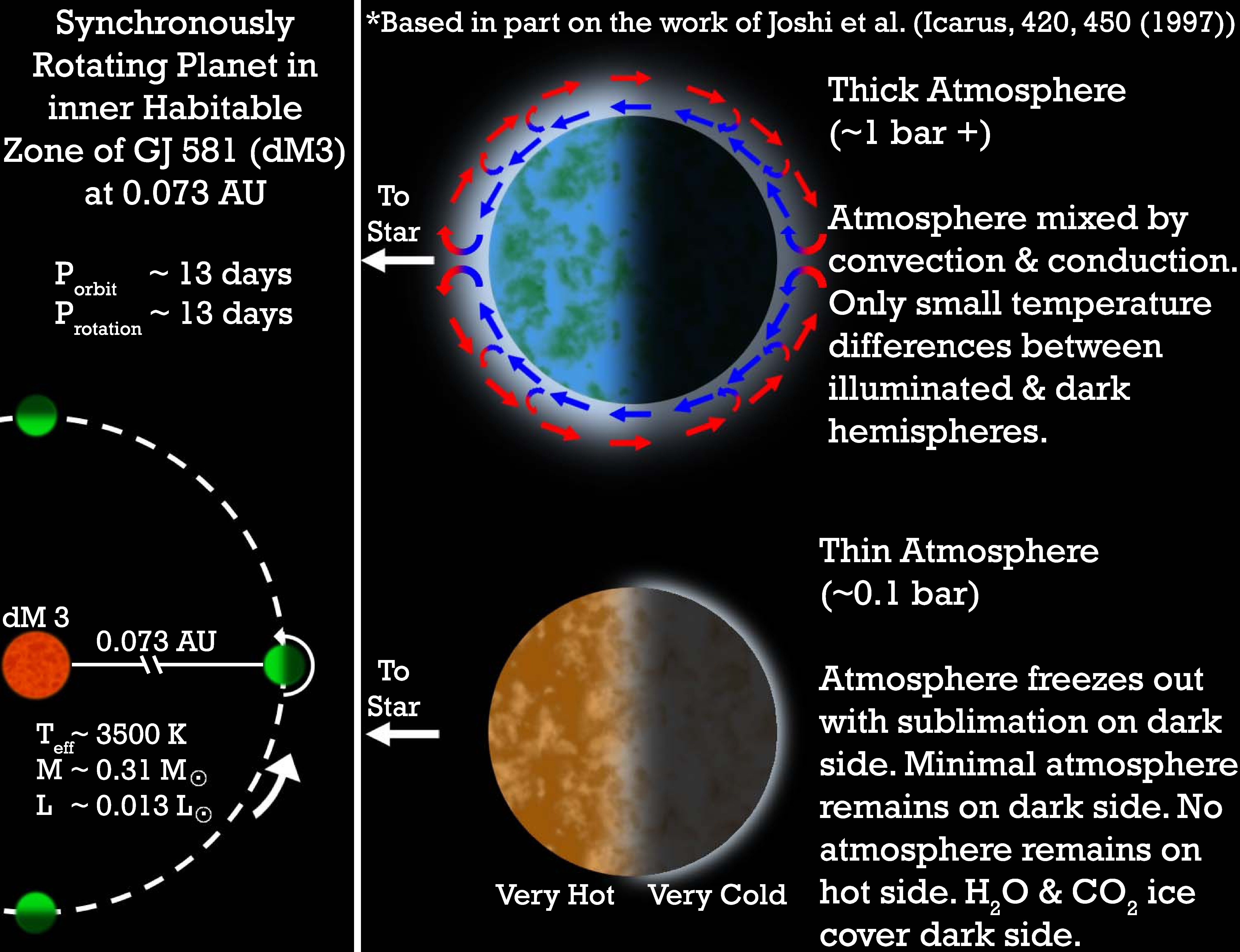}
\end{center}
\vspace{-5mm}
\caption{{\bf Left --} The currently known dM Star Planetary Systems are
given along with the properties of the host stars and planets.
{\bf Right --} Illustration of the planet GJ 581c and the effects
of either a thick or thin atmosphere.}
\vspace{-5mm}
\end{figure}

The luminosities of dM stars are essentially constant over tens to hundreds
of billions of years after their arrival on the main sequence (Fig. 1b). For
this reason, their HZs remain fixed over eons of time, ensuring a stable
energy source for hosted planets. By contrast, stars like the Sun undergo
significant changes in luminosity on timescales of a few billion years,
causing their HZs to move slowly outward over time.

\section{Extrasolar Planets Around dM Stars}

Until the recent theoretical studies of \citet{boss2006a,boss2006b}, which showed
planets can readily form around dM stars, it was thought that the low
masses of dM stars might preclude their formation, due to the expected
insufficiently massive protoplanetary disks and possible loss of planets
due to lower gravity. Moreover, this study showed that the majority of
dM star planetary systems may be composed of large terrestrial (Earth-like)
planets instead of gas giants. Except for a few studies (e.g. 
\citet{delfosse1999,vogt2000,endl2003}), dM stars have not been
specifically targeted in major extrasolar
planet searches. Nevertheless, planets now have been discovered orbiting eight
nearby dM stars using Doppler velocity techniques
(\citet{rivera2005,forveille2008}). The properties of these planets are given
in Fig. 2a. An increasing number of sub-Neptune mass ($<$ 0.1 M$_{\rm J}$)
planets and ``Super-Earths'' (defined loosely as planets with masses in the
range: 2M$_\oplus$ $<$ $M_{\rm P}~sin~i$ $<$ 10 M$_\oplus$) have been discovered to be hosted by
dM stars. As pointed out recently by \citet{forveille2008}, there is
growing observational evidence that low mass planets
($<$ 0.1M$_{\rm J}$) may be common around red dwarf stars. For example, at the
time of writing, $\sim$$\onethird$ of the 20 known planets with
$M_{\rm P}~sin~i$ $<$ 0.1 M$_{\rm J}$ and three of the seven known Super-Earths have
been found to orbit dM stars. 

Interestingly, one of these stars -- GJ 876 (IL Aqr: dM4, d = 4.7 pc) --
has been found to host a sub-Jupiter size planet orbiting within its HZ
as well as a ``Super-Earth''  with a mass of only $\sim$7.5 M$_\odot$
\citep{rivera2005}. And more recently, using the HST Advanced Camera
for Surveys (ACS), several additional dM stars may have been found to host
Jupiter-/Neptune-size planets using photometric transit methods
\citep[see][]{sahu2006}. The recent most exciting discovery is that of a
large Earth-size planet (GJ 581c: P = 13-d, M $\approx$ 5 M$_\odot$,
R $\approx$ 1.5 R$_\odot$) that orbits within the inner warm edge of
the host dM3 star's HZ. The host star, GJ 581, 
is already known to harbor a Neptune mass planet and possibly a third
planet with a mass of about 8 M$_\oplus$ \citep{udry2007}. Also see
a recent discussion of the habitability of the GJ 581 system by
\citet{selsis2007}. Because of its low luminosity, the HZ of
GJ 581 lies between $\sim$0.07--0.15 AU from the star. The discovered
planet (GJ 581c) orbits $\sim$0.07 AU from
its host star and the mean temperature on that planet should range
between 0 and 40$\deg$C allowing the presence of liquid water on its
surface \citep{udry2007}. Orbiting so close to its host star, though,
the planet will become tidally locked (P$_{\rm rot}$ = P$_{\rm orb}$).
This could initially be thought to hinder habitability. However,
\citet{joshi1997} have shown that a thick enough atmosphere (P $>$ 0.1 bar)
is capable of transfering
heat from the star-lit side of the planet to the dark hemisphere, preventing
atmospheric collapse and possibly moderating the global climate (see Fig. 2b).
Additionally, Selsis et al. conclude that the more distant planet GJ 581d
could also be habitable, given a strong enough greenhouse effect. 

\begin{figure}[!t]
\begin{center}
\includegraphics[width=0.48\textwidth]{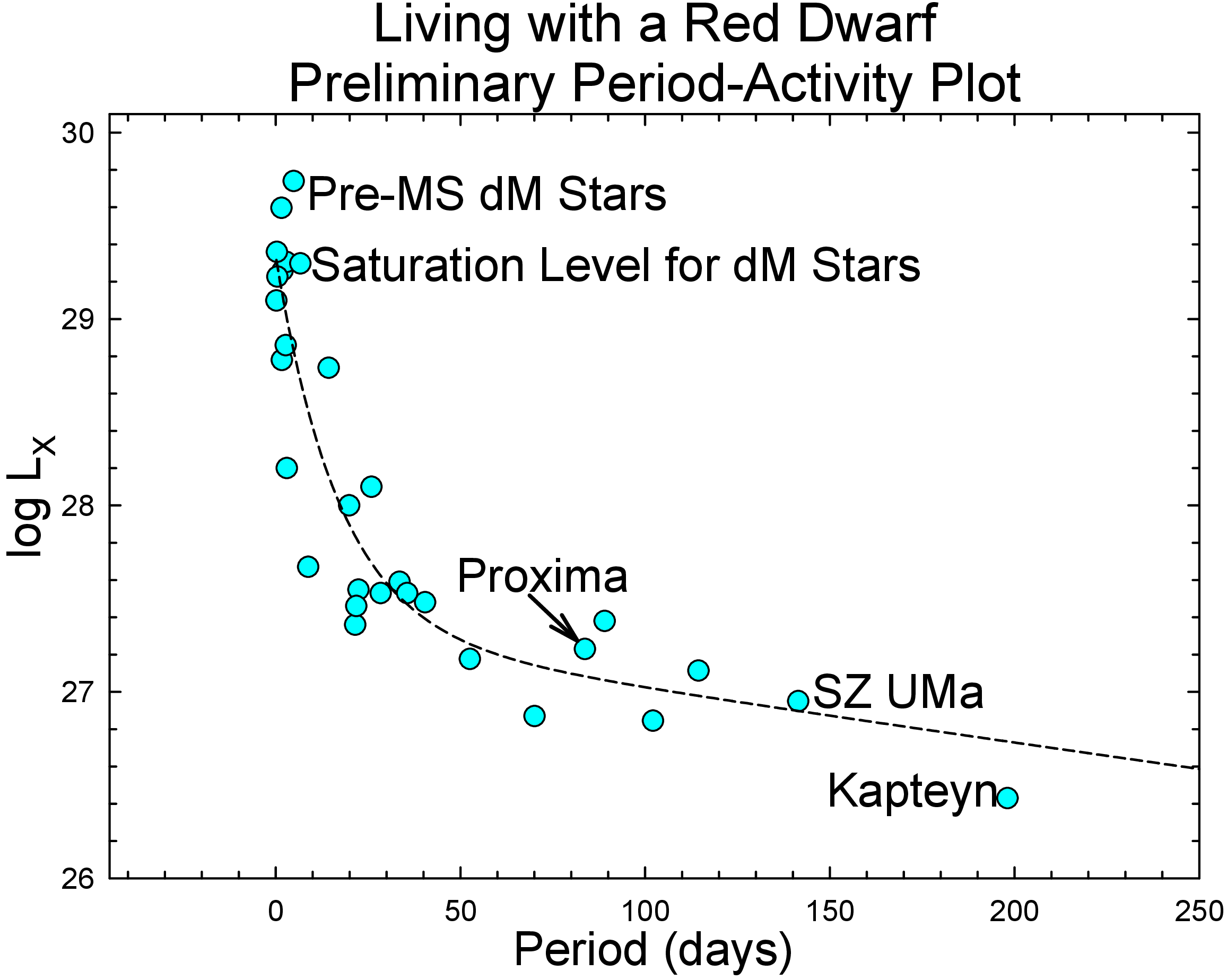}
\includegraphics[width=0.48\textwidth]{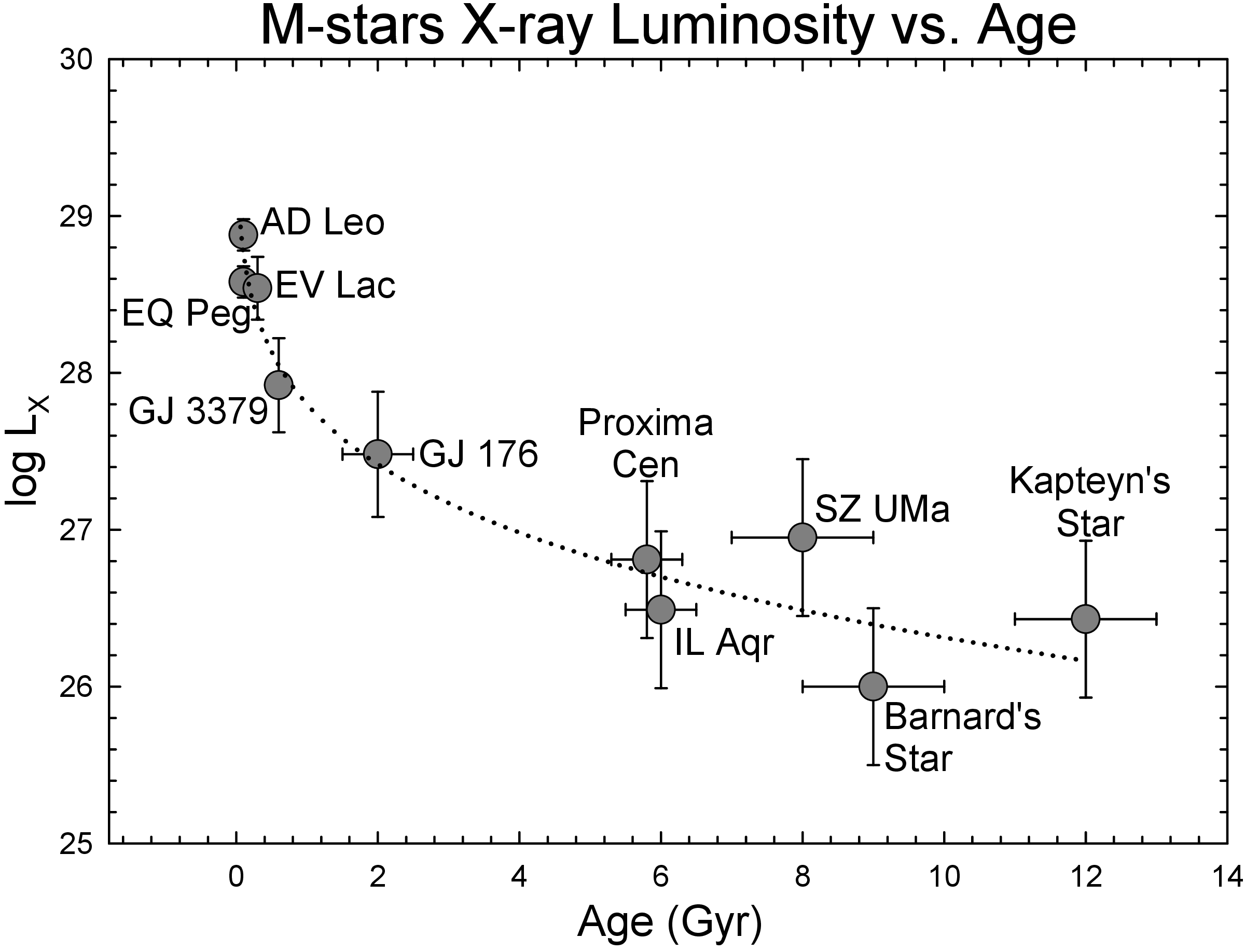}
\end{center}
\vspace{-5mm}
\caption{{\bf Left --} The coronal X-ray luminosities (log L$_{\rm x}$ in
ergs/s) are plotted against rotation period for a representative sample of
dM0--6 stars from the ``Living with a Red Dwarf'' Program. Note the rapid
decrease in coronal X-ray emission with increasing rotation period.
{\bf Right --} Log L$_{\rm x}$
values of dM stars with reliable age estimates are plotted versus age.}
\vspace{-5mm}
\end{figure}

\section{Preliminary Results of the ``Living with a Red Dwarf'' Program:
Age-Rotation-Activity Relations for dM stars}

We are carrying out high precision multi-band photometry of a small
sample of dM0--6 stars to determine rotation periods from star spot
induced light variations. As found in this study, the rotation period of a
dM star is indicative of its age. Young dM stars spin rapidly and have
correspondingly robust dynamos \& magnetic activity. Over time, though,
they lose angular momentum via magnetized stellar winds and their rotation
periods lengthen. Unfortunately, reliable ages cannot be determined for dM
stars by fitting their observable properties with theoretical evolutionary
tracks because these properties ($L$, $T_{\rm eff}$, $R$) change very slowly
over time in dM stars. However, the ages of some dM stars can be
reliably determined from: memberships in nearby star clusters or moving
groups (from common kinematic properties) that have reliable ages
(almost entirely $<$ 2 Gyr), or from sufficiently high $UVW$ space
motions which suggest either Old Disk (6--10 Gyr) or Pop II Halo ($>$10 Gyr)
ages. Another possible method of determining accurate ages for dM
stars, though, is from memberships in wide binaries or common proper
motion pairs where the more massive/luminous companion star's age is
known from fits to evolutionary tracks. For
example, the membership of the nearby 11$^{\rm th}$ mag dM5 star
Proxima Cen as an outlying member of the $\alpha$ Cen triple star
system has allowed for an accurate age determination from isochronal
fits to the G2V system member $\alpha$ Cen A. This has established a
reliable age of $\sim$5.8$\pm$0.5 Gyr for a star whose age would otherwise be
indeterminate. Also, dM stars that are paired with white dwarf companions
can have their ages inferred by the white dwarf cooling times. 
When combined with measures of X-ray, UV and H$\alpha$
emissions, accurate Age-Rotation-Activity Relationships can be constructed.

Over the last three years we have been carrying out exploratory photoelectric
photometry of a small representative sample of dM0--6 stars to reliably
determine rotation periods and star spot properties. The rotation periods are
found through low amplitude light variations arising from the presence of star
spots. Most of this $UBVRI$ (and in some cases Titanium Oxide (TiO))
photometry is being conducted using
the 0.8-m Four College Automatic Photoelectric Telescope (FCAPT) located
in Arizona. This study has uncovered low amplitude rotational light
modulations (and thus rotation periods) for many of these stars.
There is also strong evidence of long-term light variations in some of
these stars (e.g. Proxima Cen -- P$_{\rm cyc}$ $\approx$ 7 yr),
indicative of possible solar-like magnetic activity cycles.
We have also been searching for
light variations in additional equatorial and southern dM stars (and also
dK stars) included in the All Sky Automated
Survey (ASAS-3; \citet{pojmanski2001}). Utilizing the powerful period search
routines in the latest version of the Period Analysis Software
(Peranso -- http://www.peranso.com) program, strong
evidence of periodicity and possibly long-term systematic variations in
brightness have been uncovered for dozens of additional dM0--6 stars,
including Proxima Cen.

Fig. 3a,b illustrates one of the most valuable results of this
study -- the Age-Rotation-Activity relationships (example given
is coronal X-ray activity obtained from the ROSAT archives). Many of the
stars with ages $\leq$ 2 Gyr are members of clusters, moving groups
or associations whose ages have been extensively studied and reliably
determined. Age estimates for stars older than $\sim$7 Gyr have been
inferred from their large $UVW$ space motions and galactic orbits. Given
the extensive X-ray satellite archives (particularly the ROSAT All-Sky
Survey), it is hard to overestimate the value of tight, reliable
relationships between dM star X-ray activity and age. The ability to
reliably estimate the age of a dM star of known spectral type by
measuring the star's coronal L$_{\rm x}$ (or f$_{\rm x}$) value will
be of great use to future studies concerning
dM stars. Additional results from our preliminary analysis of FUV--NUV
emission lines from the extensive IUE archive have shown just as much
potential. These emissions are important for photoionization and
photochemical reactions in the upper planetary atmospheres. Also under
study are the frequent flares that dM stars display, in which their UV fluxes
increase 10--100x for several minutes. The increased X-UV radiation
from flares could have adverse effects on the retention of a planet's
atmosphere and be harmful to possible life on its surface.

\section{Conclusions \& Future Prospects}

The ``Living with a Red Dwarf'' Program, as described here, will benefit the
field of extrasolar planets by providing the ages and magnetic evolution
(and resulting X-UV emissions)
of their host stars. The Age-Rotation-Activity relationships for dM
stars found through this program will allow non-program dM stars to have
reliably determined ages based on photometric rotation periods or measures of
X-ray, UV or
H$\alpha$ emissions (and accurate spectral type for best precision).
With the ages of dM star planetary systems reliably known, the dynamic
evolutionary history and stability of the system can then be assessed.
Furthermore, the past, present and future X-UV radiative environments that the
planets will face can be estimated. Finally, planetary
habitability can be assessed based on the irradiance data and statistical
possibilities can be assigned to whether or not life could originate and
evolve on a planet hosted by a red dwarf. 

Thus, potentially habitable planets
around dM stars are clearly an important issue for future study. Dedicated
search programs for planets orbiting dM stars are being developed. For
example, the precision radial velocity spectrometer (PRVS -- 1 m/s precision
in the near-IR) is planned for use on the Gemini telescope
by 2011 \citep{jenkins2007}. Another important ground-based
program is the The MEarth Project - a transit survey of $\sim$2000 dM stars
in the northern hemisphere \citep{irwin2008}. This
project will eventually consist of eight 0.4-m robotic telescopes (two have
been in operation since December 2007) and can detect transits by bodies as
small as $\sim$2 R$_\oplus$. Also, space-based planet search missions,
including CoRoT and the upcoming Kepler, SIM PlanetQuest and Darwin/TPF
Missions are planning to target many additional dM stars. It now appears
timely to assess the likelihood of life not only around main sequence dG \&
dK stars, like our Sun, but also around the very numerous, low
luminosity dM stars.

\acknowledgements

This research is supported by NSF/RUI Grant No. AST-05-07542 and
NASA/FUSE Grant No. NNX-06AD386 which we gratefully acknowledge.


\begin{thebibliography}{}

\bibitem[Bochanski et al.(2007)]{bochan2007}{Bochanski}, J.J., {Munn}, J.A.,
{Hawley}, S.L., {West}, A.A., {Covey}, K.R. \& {Schneider}, D.P. 2007, \aj,
134, 2418

\bibitem[Boss(2006b)]{boss2006b}{Boss}, A.P. 2006, \apj, 644, 79

\bibitem[Boss(2006a)]{boss2006a}{Boss}, A.P. 2006, \apj, 643, 501

\bibitem[Delfosse et al.(1999)]{delfosse1999}{Delfosse}, X., {Forveille},
T., {Beuzit}, J.-L., {Udry}, S., {Mayor}, M. \& Perrier, C. 1999, \aap,
344, 897

\bibitem[Endl et al.(2003)]{endl2003}{Endl}, M., {Cochran}, W.D., {Tull},
R.G. \& {MacQueen}, P.J. 2003, \aj, 126, 3099

\bibitem[Forveille et al.(2008)]{forveille2008}{Forveille}, T., {Bonfils},
X., {Delfosse}, X., {Gillon}, M., {Udry}, S., {Bouchy}, F., {Lovis}, C.,
{Mayor}, M., {Pepe}, F., {Perrier}, C., {Queloz}, D., {Santos}, N. \&
{Bertaux}, J.-L. 2008, $arXiv:0809.0750$

\bibitem[Guinan et al.(2007)]{guinan2007}{Guinan}, E.F., {Engle}, S.G.,
{Ribas}, I., {Schulze-Makuch}, D. \& {McCook}, G.P. 2007, \baas, 38, 145

\bibitem[Guinan \& Engle(2007)]{guinanengle2007}{Guinan}, E.F. \&
{Engle}, S.G. 2007, $arXiv:0711.1530$

\bibitem[Irwin et al.(2008)]{irwin2008}{Irwin}, J., {Charbonneau}, D.,
{Nutzman}, P. \& {Falco}, E. 2008, $arXiv:0807.1316$

\bibitem[Jenkins et al.(2007)]{jenkins2007}{Jenkins}, J., {Jones}, H.R.A. \&
{Ramsey}, L. 2007, \baas, 39, 767

\bibitem[Joshi et al.(1997)]{joshi1997}{Joshi}, M.M., {Haberle}, R.M. \&
{Reynolds}, R.T. 1997, $Icarus$, 129, 450

\bibitem[Kasting et al.(1993)]{kasting1993}{Kasting}, J.F., {Whitmire},
D.P. \& {Reynolds}, R.T. 1993, $Icarus$, V. 101, Issue 1, p. 108

\bibitem[Kasting \& Catling(2003)]{kastingcatling2003}{Kasting}, J.F. \&
{Catling}, D. 2003, \araa, 41, 429

\bibitem[Lammer et al.(2007)]{lammer2007}{Lammer}, H., {Lichtenegger},
H.I.M., {Kulikov}, Y.N., {Grie{\ss}meier}, J.-M., {Terada}, N., {Erkaev},
N.V., {Biernat}, H.K., {Khodachenko}, M.L., {Ribas}, I., {Penz}, T. \&
{Selsis}, F. 2007, $Astrobiology$, V. 7, Issue 1, p. 185

\bibitem[Pojmansk(2001)]{pojmanski2001}{Pojma\'{n}ski}, G. 2001, $ASPC$,
246, 53

\bibitem[Reiners \& Basri(2007)]{reinersbasri2007}{Reiners}, A. \&
{Basri}, G. 2007, \apj, 656, 1121

\bibitem[Rivera et al.(2005)]{rivera2005}{Rivera}, E.J., {Lissauer}, J.J.,
{Butler}, R.P., {Marcy}, G.W., {Vogt}, S.S., {Fischer}, D.A., {Brown},
T.M., {Laughlin}, G. \& {Henry}, G.W. 2005, \apj, 634, 625

\bibitem[Sahu et al.(2006)]{sahu2006}{Sahu}, K.C., {Casertano}, S., {Bond},
H.E., {Valenti}, J., {Ed Smith}, T., {Minniti}, D., {Zoccali}, M., {Livio},
M., {Panagia}, N., {Piskunov}, N., {Brown}, T.M., {Brown}, T., {Renzini},
A., {Rich}, R.M., {Clarkson}, W. \& {Lubow}, S. 2006, \nat, V. 443, Issue
7111, p. 534

\bibitem[Scalo et al.(2007)]{scalo2007}{Scalo}, J., {Kaltenegger}, L.,
{Segura}, A.G., {Fridlund}, M., {Ribas}, I., {Kulikov}, Y.N., {Grenfell},
J.L., {Rauer}, H., {Odert}, P., {Leitzinger}, M., {Selsis}, F.,
{Khodachenko}, M.L., {Eiroa}, C., {Kasting}, J. \& {Lammer}, H. 2007,
$Astrobiology$, V. 7, Issue 1, p. 85

\bibitem[Selsis et al.(2007)]{selsis2007}{Selsis}, F., {Kasting}, J.F.,
{Levrad}, B., {Paillet}, J., {Ribas}, I. \& {Delfosse}, X. 2007, \aap,
476, 1373

\bibitem[Tarter et al(2007)]{tarter2007}{Tarter}, J.C., {Backus}, P.R.,
{Mancinelli}, R.L., {Aurnou}, J.M., {Backman}, D.E., {Basri}, G.S., {Boss},
A.P., {Clarke}, A., {Deming}, D., {Doyle}, L.R., {Feigelson}, E.D., {Freund},
F., {Grinspoon}, D.H., {Haberle}, R.M., {Hauck}, S.A., II, {Heath}, M.J.,
{Henry}, T.J., {Hollingsworth}, J.L., {Joshi}, M.M., {Kilston}, S., {Liu},
M.C., {Meikle}, E., {Reid}, I.N., {Rothschild}, L.J., {Scalo}, J., {Segura},
A., {Tang}, C.M., {Tiedje}, J.M., {Turnbull}, M.C., {Walkowicz}, L.M.,
{Weber}, A.L. \& {Young}, R.E. 2007, $Astrobiology$, V. 7, Issue 1, p. 30

\bibitem[Udry et al.(2007)]{udry2007}{Udry}, S., {Bonfils}, X., {Delfosse},
X., {Forveille}, T., {Mayor}, M., {Perrier}, C., {Bouchy}, F., {Lovis}, C.,
{Pepe}, F., {Queloz}, D. \& Bertaux, J.-L. 2007, \aap, 469, 43

\bibitem[Vogt et al.(2000)]{vogt2000}{Vogt}, S.S., {Marcy}, G.W., {Butler},
R.P. \& {Apps}, K. 2000, \apj, 536, 902






\end{thebibliography}
\end{document}